\documentclass{article}
\usepackage{spconf,amsmath,graphicx,hyperref}

\usepackage{booktabs}
\usepackage{multirow}
\usepackage{ragged2e}   
\usepackage{microtype}
\usepackage{xcolor}

\usepackage[font=small]{caption}

\title{ECHOFAKE: A REPLAY-AWARE DATASET FOR PRACTICAL SPEECH DEEPFAKE DETECTION}
\name{
    Tong Zhang\textsuperscript{\rm 1}, 
    Yihuan Huang\textsuperscript{\rm 1},
    Yanzhen Ren\textsuperscript{\rm 1,2,*}\thanks{$*$ Corresponding author. \\
This work is supported by the Natural Science Foundation of China (NSFC) under the grant NO.62572358, 62172306, 62372334.}}
\address{    \textsuperscript{\rm 1}School of Cyber Science and Engineering, Wuhan University \\
    \textsuperscript{\rm 2}Key Laboratory of Aerospace Information Security and Trusted Computing, Ministry of Education}
\begin{document}
\ninept
\maketitle
\begin{abstract}
The growing prevalence of speech deepfakes has raised serious concerns, particularly in real-world scenarios such as telephone fraud and identity theft. While many anti-spoofing systems have demonstrated promising performance on lab-generated synthetic speech, they often fail when confronted with physical replay attacks—a common and low-cost form of attack used in practical settings. Our experiments show that models trained on existing datasets exhibit severe performance degradation, with average accuracy dropping to 59.6\% when evaluated on replayed audio. To bridge this gap, we present EchoFake, a comprehensive dataset comprising more than 120 hours of audio from over 13,000 speakers, featuring both cutting-edge zero-shot text-to-speech (TTS) speech and physical replay recordings collected under varied devices and real-world environmental settings. Additionally, we evaluate three baseline detection models and show that models trained on EchoFake achieve lower average EERs across datasets, indicating better generalization. By introducing more practical challenges relevant to real-world deployment, EchoFake offers a more realistic foundation for advancing spoofing detection methods.
\end{abstract}
\section{Introduction}
\label{sec:intro}

\begin{figure}[t]
\centering
\includegraphics[width=1.0\columnwidth]{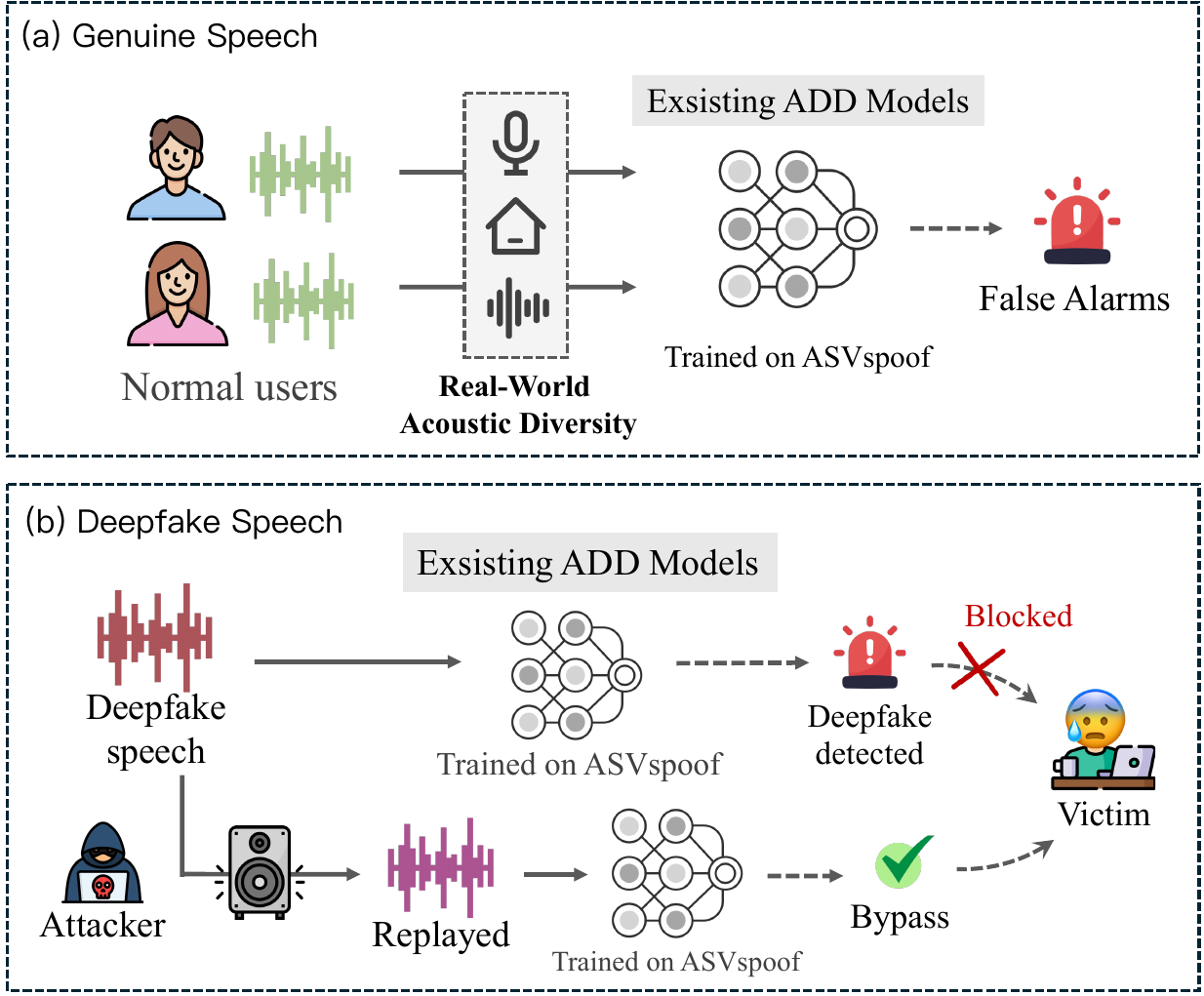} 
\caption{ADD models trained on the ASVspoof dataset often fail in real-world conditions. (a) When ordinary users employ existing ADD models, the models are prone to misclassifying genuine speech as fake due to variations in recording environments and speaker characteristics. (b) Attackers can simply replay synthetic speech to bypass detection, causing the model to misclassify fake speech as genuine. 
}
\label{fig1}
\vspace{-1ex}
\end{figure}

The recent advances in zero-shot text-to-speech (TTS) and large-scale audio language models (ALM) have dramatically lowered the barrier to generating high-quality synthetic speech. With only a few seconds of reference audio, these models can convincingly clone a speaker’s voice, producing speech that is perceptually indistinguishable from genuine utterances. While these technologies offer exciting applications in personalized voice assistants, voiceovers, and accessibility, they also pose a serious threat to security and public trust. The proliferation of voice cloning tools has opened up new avenues for audio-based forgery, impersonation, and misinformation.

While many audio deepfake detection (ADD) systems achieve strong performance on in-domain benchmarks, their robustness in real-world conditions remains a major concern. A key reason is overfitting—many ADD models are trained on clean, studio-quality datasets such as ASVspoof2019 logical access (LA) track, which leads to poor generalization when deployed in the wild. To mitigate overfitting, several datasets have been proposed to simulate more realistic conditions. ASVspoof 2021\cite{asvspoof2021} LA track focuses on telephone-channel scenarios, while ASVspoof 2021 deepfake (DF) track addresses challenges from lossy compression and encoding. The InTheWild \cite{inthewild} dataset further collects fake and genuine samples from social media platforms to increase data diversity. 

Despite recent progress, three major challenges persist for audio deepfake detection (ADD). First, models often misclassify bona fide speech recorded on consumer devices as spoofed, raising serious usability concerns (Figure~\ref{fig1}a). Second, many systems are trained on outdated spoofing methods (e.g., ASVspoof 2019 LA), failing to reflect advances in neural TTS and voice cloning. Most critically, detection remains highly vulnerable to replay attacks: by playing synthetic audio through a speaker and re-recording it, attackers can mask artifacts and deceive models into accepting it as genuine speech (Figure~\ref{fig1}b). In high-stakes applications such as telephone fraud, adversaries may even replay authentic voice snippets from prior conversations or social media, making detection especially difficult since the audio truly originates from the victim.

Existing benchmarks like ASVspoof 2019 physical access (PA) and 2021 PA tracks have laid the groundwork for replay attack detection, yet their datasets rely on software-simulated playback rather than actual physical recordings. More critically, while the PA tracks contain replayed bona fide speech, they lack any instances of replayed deepfake speech. This omission limits their relevance to emerging threats, where attackers may clone a victim’s voice and physically replay it to bypass anti-spoofing systems.

To address these challenges, we introduce EchoFake, a novel dataset that integrates zero-shot TTS deepfakes with varied physical replay conditions, offering a more realistic benchmark for spoofing detection. Using EchoFake, we reveal that existing anti-spoofing models suffer severe performance drops under diverse replay scenarios and remain vulnerable even after training, especially when distinguishing replayed bona fide speech. Nevertheless, incorporating replay diversity improves generalization: models trained on EchoFake achieve lower average EERs across multiple benchmarks, highlighting both the weaknesses of prior datasets and the benefits of realistic replay data for robust detection.
\vspace{-1ex}

\section{Related Work}
\label{sec:related_work}

\subsection{Audio Deepfake Detection}
Audio deepfake detection aims to distinguish bona fide utterances from spoofed or synthesized ones. Existing approaches can be broadly categorized into pipeline-based and end-to-end paradigms. 1) Pipeline-based detectors~\cite{lavrentyeva2019stc, Wang2021ACS, 9746688, xue2022audio} typically adopt a two-stage strategy: handcrafted or pretrained features—such as mel-frequency cepstral coefficients (MFCCs), linear frequency cepstral coefficients (LFCCs), or wav2vec2 embeddings~\cite{wav2vec2}—are first extracted, followed by a separate classifier for final prediction. This modular design allows flexible integration of domain knowledge and facilitates interpretability and transferability across datasets and spoofing conditions. 2) End-to-end methods~\cite{hua2021towards, aasist, rawnet2, xue2023learning}, on the other hand, learn task-specific representations directly from raw waveforms by jointly optimizing feature extraction and classification. These models benefit from strong representation learning capabilities and can automatically adapt to data variations without manual feature engineering. 
\vspace{-1.5ex}

\subsection{Audio Deepfake Datasets}

A variety of datasets have been developed to support research in audio deepfake detection. Classic benchmarks such as ASVspoof 2019 LA and ASVspoof 2021~\cite{asvspoof2019,asvspoof2021} focused on spoofing attacks targeting automatic speaker verification systems. The In-the-Wild~\cite{inthewild} dataset collects synthetic speech samples of celebrities from the Internet, offering a more realistic setting for deepfake detection. The ADD challenge series~\cite{add2022} and ASVspoof 5~\cite{asvspoof5} incorporates codec artifacts and multi-domain content for broader coverage. However, most datasets focus solely on either synthesis or replay. Our proposed EchoFake dataset fills this gap by integrating TTS-generated and replayed speech under diverse recording configurations, supporting more comprehensive model evaluation under realistic deployment conditions.
\vspace{-1.8ex}

\section{EchoFake Dataset}

To tackle the mismatch between lab-generated synthetic datasets and real-world spoofing scenarios involving replay attacks, we construct EchoFake, a new dataset that broadens the scope of audio deepfake detection beyond synthetic samples alone.

EchoFake consists of four subsets: training, development, closed-set evaluation, and open-set evaluation. The first three share the same pool of speakers and TTS systems to support in-distribution training and tuning, while the open-set introduces unseen speakers, new spoofing systems, and diverse replay conditions, providing a rigorous test of model generalization in realistic settings.

\subsection{Dataset Construction}

\begin{figure}[t]
\centering
\includegraphics[width=1.0\columnwidth]{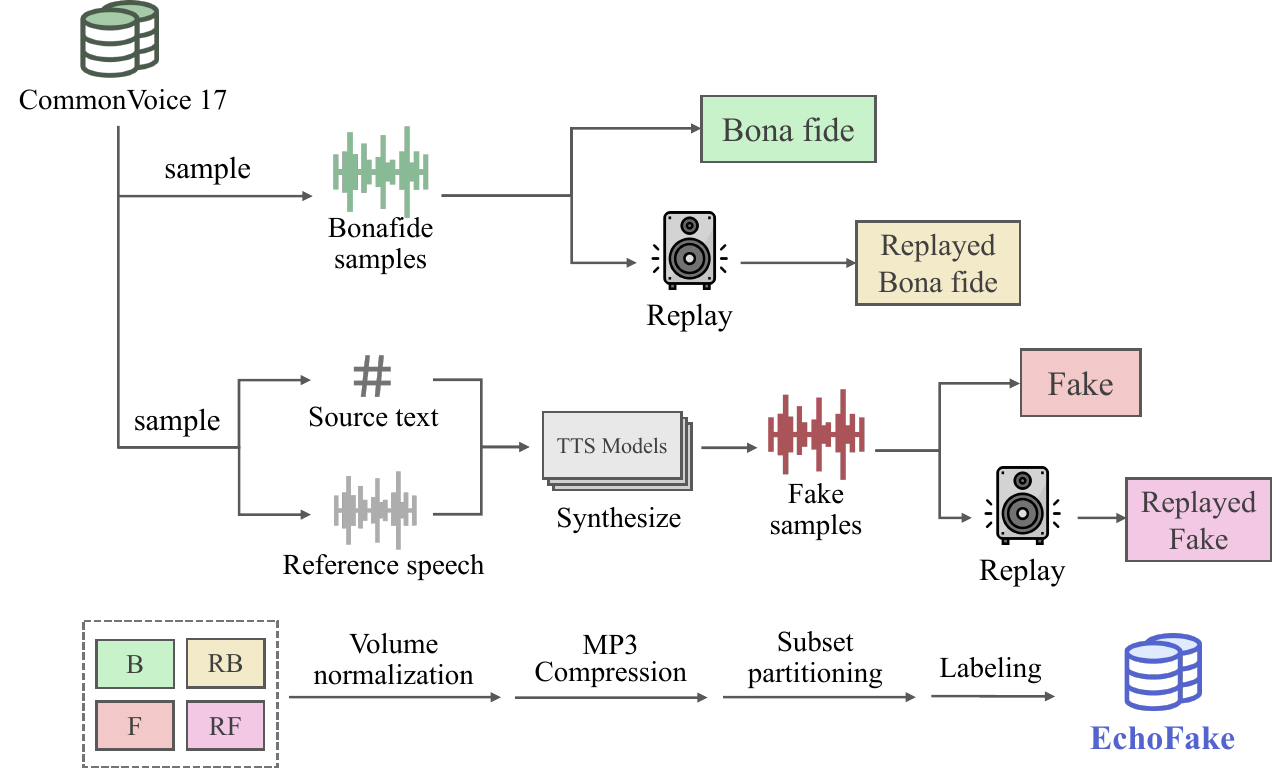} 
\caption{Pipeline for constructing the EchoFake dataset. Abbreviations: B (bona fide), RB (replayed bona fide), F (fake), RF (replayed fake). }
\label{fig2}
\vspace{-1ex}
\end{figure}



\subsubsection{Sampling and Generation}

As shown in Figure~\ref{fig2}, bona fide speech samples are directly extracted from the Common Voice 17.0 dataset. Half of these samples were replayed to construct the replayed bona fide subset. For fake speech, we randomly sampled source texts and reference speech clips from Common Voice and then use zero-shot TTS models to synthesize new utterances with the target speaker's voice cloned from the reference speech clips. Similarly, 50\% of the generated utterances are replayed to obtain the replayed fake subset. All sampling operations are strictly non-redundant to maximize diversity.

\subsubsection{TTS Model Selection}

For TTS model selection, we considered three key criteria: (1) widespread adoption or popularity in TTS communities (e.g., GitHub, HuggingFace); (2) employing state-of-the-art techniques in speech synthesis with high intelligibility, naturalness and fidelity; and (3) support for zero-shot voice cloning, which enables speaker mimicry, a common strategy in voice-based spoofing attacks.

Based on the aforementioned selection criteria, six state-of-the-art TTS models were adopted for generating fake speech across three subsets: the training set, development set, and closed-set evaluation set. Specifically, we employed: (1) \textbf{XTTSv2}~\cite{xtts}: Multilingual; VQ-VAE architecture with GPT-2 encoder and HiFi-GAN vocoder; (2) \textbf{F5-TTS}~\cite{f5-tts}: Flow matching using diffusion transformers, with sway sampling strategy; (3) \textbf{SpeechT5}~\cite{speecht5}:  Unified encoder-decoder; Multi-task pretraining; Cross-modal vector quantization; (4) \textbf{LLaSA-1B}~\cite{llasa}: Large language model (LLM) with codec-based audio adapter; (5) \textbf{OpenAudio-S1}~\cite{fishspeech}: LLM and dual autoregressive architecture, with FFGAN vocoder; (6) \textbf{StyleTTS2}~\cite{styletts2}: Style diffusion and adversarial training with speech language models.

To evaluate cross-model robustness against diverse spoofing threats, five additional models were incorporated exclusively  for open-set evaluation: (1) \textbf{CosyVoice2}~\cite{cosyvoice2}: Streaming TTS; Progressive semantic decoding with LLMs and flow matching; (2) \textbf{IndexTTS}~\cite{indextts}: Bilingual TTS; Conformer-based speech conditional encoder; BigVGAN vocoder; (3) \textbf{MaskGCT}~\cite{maskgct}: Masked generative transformers with discrete speech semantic representation; (4) \textbf{OpenVoice-V2}~\cite{openvoice}: TTS with flexible voice style control using normalizing flows; (5) \textbf{FireRedTTS-1}~\cite{fireredtts}: Semantic-aware speech tokenizer; Decoder-only transformers with flow matching.

\begin{table}[t]
\caption{Comparison of EchoFake with existing audio deepfake detection datasets. ``\#utt'' denotes the number of utterances, ``\#spk'' the number of speakers, and ``\#gen'' the number of generation methods.}
\label{tab:dataset_comparison}
\centering
\footnotesize
\begin{tabular}{lcccc}
\toprule
\textbf{Dataset} & \textbf{Year} & \textbf{\#utt} & \textbf{\#spk} & \textbf{\#gen} \\
\midrule
ASVspoof 2019 LA       & 2019 & 122,299   & 107    & 19     \\
WaveFake               & 2021 & 117,985   & 2      & 6      \\
ASVspoof 2021 LA       & 2021 & 181,566   & 67     & 13     \\
ASVspoof 2021 DF       & 2021 & 611,829   & 93     & 100+   \\
In-the-Wild            & 2022 & 31,779    & 58     & -      \\
MLAAD                  & 2024 & 82,000    & -      & 101     \\
ASVspoof 5             & 2024 & 1,211,186 & 1,922  & 28     \\
\textbf{EchoFake (Ours)} & 2025 & 81,890    & 13,005 & 11     \\
\bottomrule
\end{tabular}
\vspace{-2ex}
\end{table}

\begin{table}[t]
\caption{Data composition of EchoFake dataset. The dataset includes four categories of audio: bona fide (B), replayed bona fide (RB), fake (F), and replayed fake (RF). It is divided into two evaluation sets—closed-set (Eval-C) and open-set (Eval-O)—with corresponding durations (Dur) reported in hours.}
\label{tab:echofake_stats}
\centering
\footnotesize
\begin{tabular}{lcccccc}
\toprule
\textbf{Split} & \textbf{\#B} & \textbf{\#RB} & \textbf{\#F} & \textbf{\#RF} & \textbf{\#Total} & \textbf{Dur} \\
\midrule
Train            & 10,000 & 9,955 & 10,004 & 9,967 & 39,926 & 62.5h \\
Dev              & 1,000  &   987 &    996 &   990 &  3,973 & 6.2h  \\
Eval-C  & 1,500  & 1,498 &  1,500 & 1,493 &  5,991 & 9.4h  \\
Eval-O    & 12,800 & 6,400 &  6,400 & 6,400 & 32,000 & 48.3h \\
\midrule
Total   & 25,300 & 18,840 & 18,900 & 18,850 & 81,890 & 126.4h \\
\bottomrule
\end{tabular}
\vspace{-2ex}
\end{table}

\subsubsection{Replay Data Acquisition}

A core contribution of EchoFake is the inclusion of replayed speech, which introduces stronger distortions than lossy compression or telephone-channel encoding. Replay not only reduces fidelity through playback and recording hardware but also adds reverberation and background noise. To capture this variability, we vary playback devices, recording devices, environments, and microphone–speaker distances, yielding 16 closed-set combinations and 4 unseen open-set conditions. Data collection is automated through a WebRTC-based application that synchronizes playback and recording across LAN-connected devices, emulating real-world scenarios such as conference calls. The implementation is open-sourced under the MIT License for reproducibility. \footnotemark[1]

To construct these replay conditions, we systematically varied playback and recording devices, environments, and microphone–speaker distances. For the training, development, and closed-set evaluation subsets, replayed speech was generated using two playback devices (a MacBook Pro 14-inch 2021 and an iPad Mini, 7th generation) and two recording devices (an iPhone 13 mini and a Samsung Galaxy A54). Recordings were captured in both a meeting room (12.8 × 9.3 × 3.2 m) and a smaller home room (4.8 × 3.2 × 3.2 m), with microphone–speaker distances of 15 cm and 50 cm. These four variables yield 16 distinct replay conditions. 

For the open-set evaluation, we introduced previously unseen configurations to test model generalization. Specifically, playback was performed using Edifier MR4 powered studio monitor speakers and a Xiaomi 13 Ultra smartphone, while recording was carried out with another Xiaomi 13 Ultra or a pair of wired earbuds with a built-in microphone. Recording was conducted in a larger office room (18.6 × 13.2 × 3.2 m) at a microphone–speaker distance of 30 cm. These settings produce four additional unseen replay conditions that further increase the diversity and challenge of EchoFake.

\footnotetext[1]{Datasets, codes and automated recording tools: \url{https://github.com/EchoFake/EchoFake/}}

\subsubsection{Postprocessing}

All four subsets undergo the same preprocessing pipeline. First, volume normalization is performed using \texttt{ffmpeg}, with the integrated loudness adjusted to $-23$ LUFS, a loudness range of $7$ LU, and a true peak limit of $-2$ dBTP. Then, MP3 compression is applied with a bitrate of 64 kbps, a sampling rate of 16 kHz, and mono-channel output to maintain compatibility with the Common Voice dataset and simulate real-world audio degradation in social media platforms. 

\subsection{Database statistics}

The proposed \textit{EchoFake} dataset comprises 81,890 utterances from 13,005 speakers and covers 11 different TTS systems. A comprehensive comparison of EchoFake with established datasets is summarized in Table~\ref{tab:dataset_comparison}. While EchoFake contains fewer utterances than mainstream datasets, it outperforms them in speaker diversity, ensuring broader representation of vocal characteristics and accent variations.


EchoFake is partitioned into four subsets: \textit{train}, \textit{dev}, \textit{closed-set evaluation}, and \textit{open-set evaluation}. The distribution of utterances across these splits is detailed in Table~\ref{tab:echofake_stats}. We maintain a balanced number of samples across four data categories—\textit{bona fide}, \textit{replayed bona fide}, \textit{fake}, and \textit{replayed fake}—resulting in a total of 126.4 hours of audio data. In addition, detailed metadata is provided for each utterance, including speaker identity, reference speech information, synthesis method, and recording environment. All metadata and dataset construction scripts have been released alongside the dataset to facilitate reproducible research. \footnotemark[1]

\section{Experiments}

\begin{table*}[htbp]
\caption{EER(\%) of three baseline models trained on different datasets and evaluated on various benchmarks. The final column (w. avg.) reports the weighted average EER, computed by first averaging the results on 19LA-eval, 21LA, and 21 DF—since they share similar synthesis methods—before combining with the remaining benchmarks. }
\label{tab:eer}
\footnotesize
\centering
\begin{tabular}{c|c|ccccccc}
\toprule
\multicolumn{2}{c}{} & \multicolumn{7}{c}{\textbf{Test Set EER\%($\downarrow $)}} \\
\cmidrule(lr){3-9}
\multicolumn{1}{c}{\textbf{Model}} & \multicolumn{1}{c}{\textbf{Train set}} & \textbf{ASV19LA} & \textbf{ASV21LA} & \textbf{ASV21DF} & \textbf{In-the-Wild} & \textbf{WaveFake} & \textbf{EchoFake-open} & \textbf{w. avg.} \\

\midrule
\multirow{4}{*}{RawNet2}
& ASV19LA-train & \textbf{6.773} & \textbf{7.979} & \textbf{22.07} & 43.02 & 56.63 & \underline{46.34} & \underline{39.57} \\
& In-the-Wild & 46.58 & 48.99 & 43.88 & -- & 54.38 & 49.84 & 50.24 \\
& WaveFake & 63.07 & 59.02 & 53.74 & \underline{41.89} & -- & 56.23 & 52.25 \\
& EchoFake-train & \underline{34.14} & \underline{36.90} & \underline{37.92} & \textbf{37.52} & \textbf{34.97} & \textbf{21.13} & \textbf{32.49} \\
\midrule
\multirow{4}{*}{AASIST}
& ASV19LA-train & \textbf{0.8295} & \textbf{3.820} & \textbf{17.17} & 43.02 & 48.28 & 43.23 & \underline{35.83} \\
& In-the-Wild & 36.42 & 44.07 & \underline{37.28} & -- & \underline{35.27} & 49.27 & 41.27 \\
& WaveFake & 40.14 & 38.52 & 46.12 & \textbf{29.13} & -- & \underline{37.81} & 36.18 \\
& EchoFake-train & \underline{31.49} & \underline{32.39} & 39.25 & \underline{37.75} & \textbf{33.59} & \textbf{14.88} & \textbf{30.15} \\
\midrule
\multirow{4}{*}{Wav2Vec2}
& ASV19LA-train & \textbf{11.69} & \textbf{8.673} & \textbf{5.244} & \textbf{11.20} & \textbf{10.31} & \underline{42.94} & \underline{18.25} \\
& In-the-Wild & \underline{11.79} & 24.06 & \underline{15.87} & -- & \underline{16.16} & 51.22 & 28.21 \\
& WaveFake & 63.78 & 64.01 & 56.69 & 49.37 & -- & 55.52 & 55.46 \\
& EchoFake-train & 14.75 & \underline{13.56} & 17.31 & \underline{16.66} & 23.41 & \textbf{11.86} & \textbf{16.79} \\
\bottomrule
\end{tabular}
\end{table*}

\subsection{Experimental Setup}

We evaluate the proposed dataset using three representative baseline systems: RawNet2, AASIST, and Wav2Vec2. For RawNet2, the model is trained for 100 epochs with a batch size of 64, an initial learning rate of $10^{-4}$. For AASIST, the model is trained for 60 epochs with a batch size of 32, an initial learning rate of $10^{-4}$. For wav2vec2, we append a multi-layer perceptron (MLP) as backend classifier and train the model in an end-to-end manner for 20 epochs with a batch size of 32, an initial learning rate of $10^{-5}$. All models are optimized using the Adam optimizer ($\beta_1=0.9$, $\beta_2=0.999$) with a weight decay of $10^{-4}$. For all models, we select the checkpoint with the best performance on the development set for final evaluation. All experiments were conducted on a single NVIDIA RTX 4090 GPU.

We evaluate detection performance using F1-score for four-class classification (B/RB/F/RF), and equal error rate (EER) and accuracy (ACC) for binary spoofing detection (spoof/bonafide). The F1-score provides balanced accuracy measurement, while EER optimizes the trade-off between false acceptance and rejection rates.



\subsection{Detection Performance on EchoFake}

To evaluate model robustness under realistic spoofing scenarios, we examine detection performance on the EchoFake dataset under two classification settings: a four-class detection setup and a conventional binary classification task.

In both four-class and binary detection tasks, models achieve strong performance in closed-set conditions but degrade sharply in open-set scenarios, with replayed samples being the main source of errors. Replayed bona fide (RB) speech is particularly difficult to detect due to its lack of synthetic artifacts and close resemblance to genuine speech, while replayed fake (RF) samples further complicate classification by combining synthesis traces with channel distortions. In contrast, purely synthetic speech remains the easiest to identify, suggesting that existing models overly rely on spectral artifacts. These results reveal a critical weakness of current anti-spoofing systems: high-fidelity replay attacks can effectively mask or distort discriminative cues, leading to substantial misclassification and undermining robustness in real-world conditions.

\begin{table}[t]
\caption{F1-score (\%) for four-class classification (top) and EER/ACC(\%) for binary classification on EchoFake test sets (bottom). Performance breakdown (F1/ACC) across all four classes is included.}
\label{tab:echo_two_class}
\centering
\setlength{\tabcolsep}{4.5pt}
\footnotesize
\begin{tabular}{ccccccc}
\toprule

\multicolumn{2}{c}{} & \multicolumn{5}{c}{\textbf{F1-score\%($\uparrow $)}} \\
\cmidrule(lr){3-7}
\multicolumn{1}{c}{\textbf{Model}} & \multicolumn{1}{c}{\textbf{Cond.}} & \textbf{Avg. F1} & \textbf{B} & \textbf{RB} & \textbf{F} & \textbf{RF} \\

\midrule
\multirow{2}{*}{RawNet2} & Closed & 94.06 & 93.89 & 94.27 & 93.94 & 94.13 \\
                         & Open   & 53.61 & 73.23 & 27.08 & 72.79 & 41.35 \\
\midrule
\multirow{2}{*}{AASIST}  & Closed & 97.63 & 99.03 & 96.33 & 98.93 & 96.21 \\
                         & Open   & 51.07 & 70.83 & 26.30 & 79.90 & 27.23 \\
\midrule
\multirow{2}{*}{Wav2Vec2} & Closed & 98.81 & 99.63 & 98.16 & 99.43 & 98.02 \\
                          & Open   & 60.99 & 78.83 & 40.13 & 75.64 & 49.34 \\
\midrule

& & \\\\ [-4ex]
\midrule
\multicolumn{2}{c}{} & \textbf{EER\%($\downarrow$)} & \multicolumn{4}{c}{\textbf{ACC\%($\uparrow$)}} \\
\cmidrule(lr){3-3} \cmidrule(lr){4-7}
\multicolumn{1}{c}{\textbf{Model}} & \multicolumn{1}{c}{\textbf{Cond.}} & \textbf{All} & \textbf{B} & \textbf{RB} & \textbf{F} & \textbf{RF} \\
\midrule
\multirow{2}{*}{RawNet2} & Closed & 3.95  & 96.07  & 99.80  & 88.40  & 99.87 \\
                         & Open   & 21.13 & 78.88  & 65.89  & 94.30  & 76.42 \\
\midrule
\multirow{2}{*}{AASIST}  & Closed & 0.46  & 99.60  & 99.93  & 98.73  & 100.00 \\
                         & Open   & 14.88 & 85.13  & 66.92  & 98.66  & 89.78 \\
\midrule
\multirow{2}{*}{Wav2Vec2} & Closed & 0.27  & 99.73  & 99.80  & 99.40  & 99.93 \\
                          & Open   & 11.86 & 88.16  & 67.64  & 99.66  & 97.13 \\
\bottomrule
\end{tabular}
\vspace{-1.8ex}
\end{table}

\subsection{Cross-Dataset Generalization}

To assess cross-dataset generalization capabilities, we conducted systematic evaluations of spoofing detection systems across multiple benchmarks. In this experiment, the RB samples from EchoFake is treated as spoofed audio, alongside F and RF samples, to align with the binary classification setting of existing datasets.



As shown in Table~\ref{tab:eer}, EchoFake exposes a critical vulnerability of current ADD models to realistic replay attacks, with all three baseline models exhibiting significant performance degradation on the open-set evaluation subset (average EER: 48\%). This deterioration highlights the challenge of developing generalizable detectors under diverse attack types and realistic acoustic conditions. Notably, models achieve their best cross-dataset performance when trained on EchoFake, suggesting its enhanced diversity in spoofing scenarios enables more effective generalization. These results establish EchoFake as a rigorous benchmark and emphasize the value of training sets that explicitly model real-world heterogeneity.

\vspace{-1ex}

\subsection{Ablation Study: Impact of Replayed Audio}

To examine the impact of replay data, we retrained all baseline models on the EchoFake training set after removing replay samples and compared their performance with models trained on the full dataset (For fair comparison, we downsampled the original training set to match the replay-excluded set's sample size). When evaluated across five conventional benchmarks (ASV19LA, ASV21LA, ASV21DF, In-the-Wild, and WaveFake), the EER changes after removing replay data were marginal: $-0.03\%$ (RawNet), $+1.67\%$ (AASIST), and $+4.07\%$ (wav2vec2). However, when including the EchoFake-open set for evaluation (six datasets in total), average EERs increased by $+1.54\%$, $+5.26\%$, and $+6.22\%$, respectively. These results demonstrate that incorporating replay data in training provides a favorable trade-off: it does not significantly degrade performance on conventional benchmarks while substantially enhancing robustness against replay-based attacks.

\vspace{-1ex}

\section{Conclusion}

In this work, we present EchoFake, a novel and comprehensive dataset designed to advance ADD system development under realistic conditions. By integrating both zero-shot TTS speech and diverse physical replay recordings, EchoFake captures spoofing patterns overlooked in existing datasets. Evaluations on EchoFake reveal that current models suffer significant performance degradation in realistic replay scenarios, highlighting critical weaknesses in both model robustness and dataset coverage. Meanwhile, incorporating EchoFake during training improves generalization across multiple benchmarks, suggesting the benefit of modeling real-world attack variability. We hope that EchoFake will serve as a valuable benchmark and contribute toward building more resilient and deployable anti-spoofing systems.

\bibliographystyle{IEEEbib}
\bibliography{strings,refs}

\end{document}